\documentclass[prl,twocolumn]{revtex4-1}

\usepackage{color} 
\usepackage{graphicx} 
\usepackage{amsmath,amssymb} 

\begin{document}

\author{Roman Shevchuk } 

\author{Francesco Rao \footnote{francesco.rao@frias.uni-freiburg.de}}
\affiliation{ Freiburg Institute for Advanced Studies, University of Freiburg,
Freiburg, Germany.}

\title{Microsecond long atomistic simulation of supercooled water}
\maketitle

Supercooled water is a metastable phase of liquid water below the melting
temperature \cite{Mishima1998}.  In this regime, the transition to the solid
phase is irreversible once the process is activated.  An interesting discussion
recently developed on the relationship between crystallization rate and the
time scales of equilibration within the liquid phase
\cite{Chandler2011,MooreMolinero2011}.  Calculations using a coarse grained monatomic
model of water, the mW model, suggested that equilibration of the liquid below
the temperature of homogeneous nucleation $T_H\approx225$~K is slower than ice
nucleation \cite{MooreMolinero2011}.  This observation has important
consequences to a proposed theory of water anomalies, predicting a second
critical point below $T_H$ where a liquid-liquid phase transition occurs
\cite{Poole1992}. Although it has attracted attention
\cite{Debenedetti2009,Abascal2010,Cuthbertson2011,Wikfeldt2011},  this theory is not
without problems. If the speed of ice nucleation is faster than liquid
relaxation, the liquid-liquid transition would loose sense from a
thermodynamical point of view, being the liquid phase not equilibrated
\cite{Chandler2011}.  

Here, a 3 $\mu$s long molecular dynamics simulation of the TIP4P-Ew water model
is presented to investigate the relaxation properties of an atomistic model in
the  supercooled region below $T_H$.  The length of this calculation is one
order of magnitude larger than the 350~ns used to study freezing with the mW
model \cite{MooreMolinero2011}. A box of 1024 molecules was simulated with
GROMACS \cite{gromacs}.  The Berendsen barostat \cite{Berendsen1984}, velocity
rescale thermostat \cite{Bussi2007} and PME \cite{Darden1993} were used for
pressure coupling, temperature coupling and long-range electrostatics,
respectively.  The simulation was run at 190~K and 1250~atm. These
values are close to the estimated liquid-liquid critical point for the TIP4P-Ew
\cite{Paschek2008}, congruous with recent calculations on the similar
TIP4P/2005 model \cite{Abascal2010}.

\begin{figure}
  \includegraphics[width=85mm]{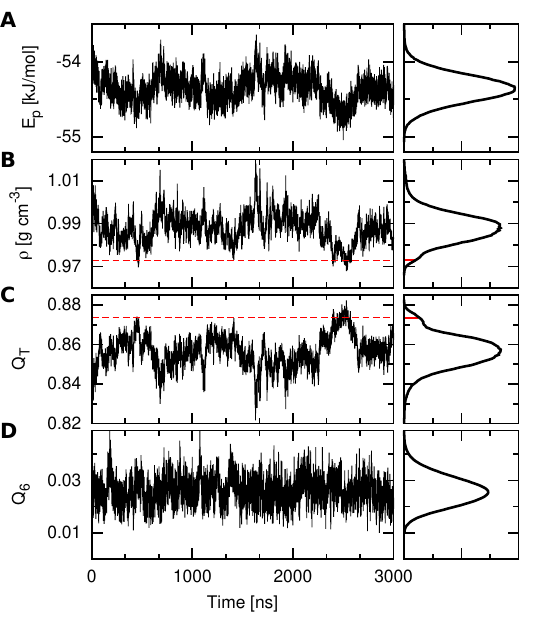}
  \caption{Time series for the 3 $\mu$s trajectory.  (A) potential
    energy; (B) density; (C) tetrahedral order parameter $Q_T$; (D) $Q_6$ parameter
    (calculated as in Ref.~\cite{Chandler2011,Sciortino2011}). Right panels show the
    probability distribution of the respective quantities. Dashed lines
    represent deviations from the mean fluctuations, with values of 0.972 and 0.873
    for $\rho$ and $Q_T$, respectively.}
  \label{fig1}
\end{figure}

In these conditions, freezing was not observed as shown by the timeseries
of the potential energy $E_p$ (Fig.~\ref{fig1}A).  Fluctuations are of the
order of 0.5 kJ/mol per molecule with no systematic drift. It has been observed
that once freezing is activated the energy drifts very quickly to low values of
the potential energy, with  large  energy changes (e.g. roughly 5 and 2 kJ/mol
per molecule for TIP4P at 230~K \cite{Ohmine2002} and TIP4P/2005 at 242~K
\cite{Fernandez2006}, respectively).

The time series of the density $\rho$ and the tetrahedral order parameter $Q_T$
\cite{Errington2001} are shown in Fig.~\ref{fig1}B-C. They respectively
correlate and anticorrelate with the potential energy (Pearson correlation coefficient $r=0.69$ and -0.86).  The distributions of
both $\rho$ and $Q_T$ show an appreciable bump at one of the tails (see right
panel of Fig.~\ref{fig1}B-C), suggesting the presence of a subpopulation. For
the case of the tetrahedral order parameter, the subpopulation emerges at
values around 0.873 (red dashed line and right side of Fig.~\ref{fig1}C). This
fluctuation is localized in a time window between 2.3 and 2.6 $\mu$s in
correspondence to a decreasing of both the density and the potential energy.
It is interesting to note that density subpopulations have been interpreted by
some \cite{Kesselring2011} as a signature of the aforementioned liquid-liquid
transition.

To check whether this fluctuation corresponded to an ice nucleation attempt,
the $Q_6$ order parameter \cite{Steinhardt1983,Chandler2011,Sciortino2011} was
calculated (Fig.~\ref{fig1}D). In the time window between 2.3-2.6 $\mu$s the
value of the parameter is around 0.025, with no signs of ice nucleation.
Moreover, no correlation with the energy was found ($r=10^{-6}$). With a value
of $Q_6$ for hexagonal ice expected to be one order of magnitude larger
\cite{Sciortino2011}, no evidence for ice nucleation is found in the present
trajectory.

\begin{figure}
  \includegraphics[width=70mm]{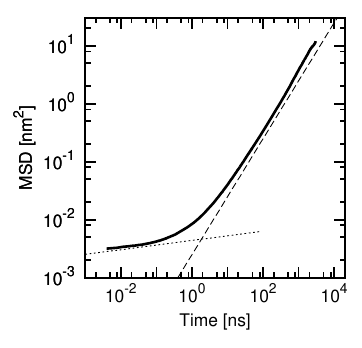}
  \caption{Oxygen mean square displacement (MSD). The dashed and dotted lines
    represent a linear and a power-law (exponent equal to 0.1) regression,
    respectively. The diffusion coefficient extracted from the linear regime is
    of $6.6 \times 10^{-9}$cm$^2$/s. The g\_msd function of GROMACS was used with 150
    windows to improve statistics.}
  \label{fig2}
\end{figure}

Finally, the oxygen mean-square-displacement (MSD) was calculated
(Fig.~\ref{fig2}). At timescales shorter than one~ns, water shows a
subdiffusive behavior (dotted line in Fig.~\ref{fig2}). For larger times the
system enters a diffusive regime, following the linear relationship MSD
$\approx t$ (dashed line), with a maximum average displacement  of 3.47 nm
after 3 $\mu$s. Taking into account that the molecular diameter is around 0.3
nm, water molecules have diffused for about 11.5 molecular diameters (the
average box side length is of 3.14 nm).

In conclusion, evidence is provided that the liquid phase of the TIP4P-Ew model
is at equilibrium in the supercooled regime before ice nucleation. Our finding
is in agreement with another $\mu$s long simulation of supercooled water with a
5-site model \cite{Kesselring2011}, suggesting that equilibration of the liquid
phase below $T_H$ is a common feature of atomistic models. The mW model has
shown to reproduce several properties of water, including density and phase
diagram \cite{Molinero2009}.  But the lack of hydrogens, and consequently of
molecular reorientations \cite{Laage2006}, might considerably speed up the time
scales.  We speculate that the differences in the relaxation kinetics between
atomistic models and the mW model are due to the lack of molecular
reorientations in the latter. Clearly, further experimental validation is
needed to clarify which proposed mechanism (if any) is closer to real water.

\vspace{1mm}
This work is supported by the Excellence Initiative of the German Federal and
State Governments.

\begin{thebibliography}{21}%
\makeatletter
\providecommand \@ifxundefined [1]{%
 \@ifx{#1\undefined}
}%
\providecommand \@ifnum [1]{%
 \ifnum #1\expandafter \@firstoftwo
 \else \expandafter \@secondoftwo
 \fi
}%
\providecommand \@ifx [1]{%
 \ifx #1\expandafter \@firstoftwo
 \else \expandafter \@secondoftwo
 \fi
}%
\providecommand \natexlab [1]{#1}%
\providecommand \enquote  [1]{``#1''}%
\providecommand \bibnamefont  [1]{#1}%
\providecommand \bibfnamefont [1]{#1}%
\providecommand \citenamefont [1]{#1}%
\providecommand \href@noop [0]{\@secondoftwo}%
\providecommand \href [0]{\begingroup \@sanitize@url \@href}%
\providecommand \@href[1]{\@@startlink{#1}\@@href}%
\providecommand \@@href[1]{\endgroup#1\@@endlink}%
\providecommand \@sanitize@url [0]{\catcode `\\12\catcode `\$12\catcode
  `\&12\catcode `\#12\catcode `\^12\catcode `\_12\catcode `\%12\relax}%
\providecommand \@@startlink[1]{}%
\providecommand \@@endlink[0]{}%
\providecommand \url  [0]{\begingroup\@sanitize@url \@url }%
\providecommand \@url [1]{\endgroup\@href {#1}{\urlprefix }}%
\providecommand \urlprefix  [0]{URL }%
\providecommand \Eprint [0]{\href }%
\providecommand \doibase [0]{http://dx.doi.org/}%
\providecommand \selectlanguage [0]{\@gobble}%
\providecommand \bibinfo  [0]{\@secondoftwo}%
\providecommand \bibfield  [0]{\@secondoftwo}%
\providecommand \translation [1]{[#1]}%
\providecommand \BibitemOpen [0]{}%
\providecommand \bibitemStop [0]{}%
\providecommand \bibitemNoStop [0]{.\EOS\space}%
\providecommand \EOS [0]{\spacefactor3000\relax}%
\providecommand \BibitemShut  [1]{\csname bibitem#1\endcsname}%
\let\auto@bib@innerbib\@empty
\bibitem [{\citenamefont {Mishima}\ and\ \citenamefont
  {Stanley}(1998)}]{Mishima1998}%
  \BibitemOpen
  \bibfield  {author} {\bibinfo {author} {\bibfnamefont {O.}~\bibnamefont
  {Mishima}}\ and\ \bibinfo {author} {\bibfnamefont {E.~H.}\ \bibnamefont
  {Stanley}},\ }\href@noop {} {\bibfield  {journal} {\bibinfo  {journal}
  {Nature}\ }\textbf {\bibinfo {volume} {396}},\ \bibinfo {pages} {329}
  (\bibinfo {year} {1998})}\BibitemShut {NoStop}%
\bibitem [{\citenamefont {Limmer}\ and\ \citenamefont
  {Chandler}(2011)}]{Chandler2011}%
  \BibitemOpen
  \bibfield  {author} {\bibinfo {author} {\bibfnamefont {D.~T.}\ \bibnamefont
  {Limmer}}\ and\ \bibinfo {author} {\bibfnamefont {D.}~\bibnamefont
  {Chandler}},\ }\href@noop {} {\bibfield  {journal} {\bibinfo  {journal} {J.
  Chem. Phys.}\ }\textbf {\bibinfo {volume} {135}},\ \bibinfo {pages} {134503}
  (\bibinfo {year} {2011})}\BibitemShut {NoStop}%
\bibitem [{\citenamefont {Moore}\ and\ \citenamefont
  {Molinero}(2011)}]{MooreMolinero2011}%
  \BibitemOpen
  \bibfield  {author} {\bibinfo {author} {\bibfnamefont {E.~B.}\ \bibnamefont
  {Moore}}\ and\ \bibinfo {author} {\bibfnamefont {V.}~\bibnamefont
  {Molinero}},\ }\href@noop {} {\bibfield  {journal} {\bibinfo  {journal}
  {Nature}\ }\textbf {\bibinfo {volume} {479}},\ \bibinfo {pages} {506}
  (\bibinfo {year} {2011})}\BibitemShut {NoStop}%
\bibitem [{\citenamefont {Poole}\ \emph {et~al.}(1992)\citenamefont {Poole},
  \citenamefont {Sciortino}, \citenamefont {Essmann},\ and\ \citenamefont
  {Stanley}}]{Poole1992}%
  \BibitemOpen
  \bibfield  {author} {\bibinfo {author} {\bibfnamefont {P.~H.}\ \bibnamefont
  {Poole}}, \bibinfo {author} {\bibfnamefont {F.}~\bibnamefont {Sciortino}},
  \bibinfo {author} {\bibfnamefont {U.}~\bibnamefont {Essmann}}, \ and\
  \bibinfo {author} {\bibfnamefont {H.~E.}\ \bibnamefont {Stanley}},\
  }\href@noop {} {\bibfield  {journal} {\bibinfo  {journal} {Nature}\ }\textbf
  {\bibinfo {volume} {360}},\ \bibinfo {pages} {324} (\bibinfo {year}
  {1992})}\BibitemShut {NoStop}%
\bibitem [{\citenamefont {Liu}\ \emph {et~al.}(2009)\citenamefont {Liu},
  \citenamefont {Panagiotopoulos},\ and\ \citenamefont
  {Debenedetti}}]{Debenedetti2009}%
  \BibitemOpen
  \bibfield  {author} {\bibinfo {author} {\bibfnamefont {Y.}~\bibnamefont
  {Liu}}, \bibinfo {author} {\bibfnamefont {A.~Z.}\ \bibnamefont
  {Panagiotopoulos}}, \ and\ \bibinfo {author} {\bibfnamefont {P.~G.}\
  \bibnamefont {Debenedetti}},\ }\href@noop {} {\bibfield  {journal} {\bibinfo
  {journal} {J. Chem. Phys.}\ }\textbf {\bibinfo {volume} {131}},\ \bibinfo
  {pages} {104508} (\bibinfo {year} {2009})}\BibitemShut {NoStop}%
\bibitem [{\citenamefont {Abascal}\ and\ \citenamefont
  {Vega}(2010)}]{Abascal2010}%
  \BibitemOpen
  \bibfield  {author} {\bibinfo {author} {\bibfnamefont {J.}~\bibnamefont
  {Abascal}}\ and\ \bibinfo {author} {\bibfnamefont {C.}~\bibnamefont {Vega}},\
  }\href@noop {} {\bibfield  {journal} {\bibinfo  {journal} {J. Chem. Phys.}\
  }\textbf {\bibinfo {volume} {133}},\ \bibinfo {pages} {234502} (\bibinfo
  {year} {2010})}\BibitemShut {NoStop}%
\bibitem [{\citenamefont {Cuthbertson}\ and\ \citenamefont
  {Poole}(2011)}]{Cuthbertson2011}%
  \BibitemOpen
  \bibfield  {author} {\bibinfo {author} {\bibfnamefont {M.}~\bibnamefont
  {Cuthbertson}}\ and\ \bibinfo {author} {\bibfnamefont {P.}~\bibnamefont
  {Poole}},\ }\href@noop {} {\bibfield  {journal} {\bibinfo  {journal} {Phys.
  Rev. Lett.}\ }\textbf {\bibinfo {volume} {106}},\ \bibinfo {pages} {115706}
  (\bibinfo {year} {2011})}\BibitemShut {NoStop}%
\bibitem [{\citenamefont {Wikfeldt}\ \emph {et~al.}(2011)\citenamefont
  {Wikfeldt}, \citenamefont {Huang}, \citenamefont {Nilsson},\ and\
  \citenamefont {Pettersson}}]{Wikfeldt2011}%
  \BibitemOpen
  \bibfield  {author} {\bibinfo {author} {\bibfnamefont {K.}~\bibnamefont
  {Wikfeldt}}, \bibinfo {author} {\bibfnamefont {C.}~\bibnamefont {Huang}},
  \bibinfo {author} {\bibfnamefont {A.}~\bibnamefont {Nilsson}}, \ and\
  \bibinfo {author} {\bibfnamefont {L.}~\bibnamefont {Pettersson}},\
  }\href@noop {} {\bibfield  {journal} {\bibinfo  {journal} {The Journal of
  Chemical Physics}\ }\textbf {\bibinfo {volume} {134}},\ \bibinfo {pages}
  {214506} (\bibinfo {year} {2011})}\BibitemShut {NoStop}%
\bibitem [{\citenamefont {Van Der~Spoel}\ \emph {et~al.}(2005)\citenamefont
  {Van Der~Spoel}, \citenamefont {Lindahl}, \citenamefont {Hess}, \citenamefont
  {Groenhof}, \citenamefont {Mark},\ and\ \citenamefont {Berendsen}}]{gromacs}%
  \BibitemOpen
  \bibfield  {author} {\bibinfo {author} {\bibfnamefont {D.}~\bibnamefont {Van
  Der~Spoel}}, \bibinfo {author} {\bibfnamefont {E.}~\bibnamefont {Lindahl}},
  \bibinfo {author} {\bibfnamefont {B.}~\bibnamefont {Hess}}, \bibinfo {author}
  {\bibfnamefont {G.}~\bibnamefont {Groenhof}}, \bibinfo {author}
  {\bibfnamefont {A.~E.}\ \bibnamefont {Mark}}, \ and\ \bibinfo {author}
  {\bibfnamefont {H.~J.~C.}\ \bibnamefont {Berendsen}},\ }\href@noop {}
  {\bibfield  {journal} {\bibinfo  {journal} {J. Comput. Chem.}\ }\textbf
  {\bibinfo {volume} {26}},\ \bibinfo {pages} {1701} (\bibinfo {year}
  {2005})}\BibitemShut {NoStop}%
\bibitem [{\citenamefont {Berendsen}\ \emph {et~al.}(1984)\citenamefont
  {Berendsen}, \citenamefont {Postma}, \citenamefont {van Gunsteren},
  \citenamefont {DiNola},\ and\ \citenamefont {Haak}}]{Berendsen1984}%
  \BibitemOpen
  \bibfield  {author} {\bibinfo {author} {\bibfnamefont {H.~J.~C.}\
  \bibnamefont {Berendsen}}, \bibinfo {author} {\bibfnamefont {J.~P.~M.}\
  \bibnamefont {Postma}}, \bibinfo {author} {\bibfnamefont {W.~F.}\
  \bibnamefont {van Gunsteren}}, \bibinfo {author} {\bibfnamefont
  {A.}~\bibnamefont {DiNola}}, \ and\ \bibinfo {author} {\bibfnamefont {J.~R.}\
  \bibnamefont {Haak}},\ }\href@noop {} {\bibfield  {journal} {\bibinfo
  {journal} {J. Chem. Phys.}\ }\textbf {\bibinfo {volume} {81}},\ \bibinfo
  {pages} {3684} (\bibinfo {year} {1984})}\BibitemShut {NoStop}%
\bibitem [{\citenamefont {Bussi}\ \emph {et~al.}(2007)\citenamefont {Bussi},
  \citenamefont {Donadio},\ and\ \citenamefont {Parrinello}}]{Bussi2007}%
  \BibitemOpen
  \bibfield  {author} {\bibinfo {author} {\bibfnamefont {G.}~\bibnamefont
  {Bussi}}, \bibinfo {author} {\bibfnamefont {D.}~\bibnamefont {Donadio}}, \
  and\ \bibinfo {author} {\bibfnamefont {M.}~\bibnamefont {Parrinello}},\
  }\href@noop {} {\bibfield  {journal} {\bibinfo  {journal} {J. Chem. Phys.}\
  }\textbf {\bibinfo {volume} {126}},\ \bibinfo {pages} {014101} (\bibinfo
  {year} {2007})}\BibitemShut {NoStop}%
\bibitem [{\citenamefont {Darden}\ \emph {et~al.}(1993)\citenamefont {Darden},
  \citenamefont {York},\ and\ \citenamefont {Pedersen}}]{Darden1993}%
  \BibitemOpen
  \bibfield  {author} {\bibinfo {author} {\bibfnamefont {T.}~\bibnamefont
  {Darden}}, \bibinfo {author} {\bibfnamefont {D.}~\bibnamefont {York}}, \ and\
  \bibinfo {author} {\bibfnamefont {L.}~\bibnamefont {Pedersen}},\ }\href@noop
  {} {\bibfield  {journal} {\bibinfo  {journal} {J. Chem. Phys.}\ }\textbf
  {\bibinfo {volume} {98}},\ \bibinfo {pages} {10089} (\bibinfo {year}
  {1993})}\BibitemShut {NoStop}%
\bibitem [{\citenamefont {Paschek}\ \emph {et~al.}(2008)\citenamefont
  {Paschek}, \citenamefont {R\"{u}ppert},\ and\ \citenamefont
  {Geiger}}]{Paschek2008}%
  \BibitemOpen
  \bibfield  {author} {\bibinfo {author} {\bibfnamefont {D.}~\bibnamefont
  {Paschek}}, \bibinfo {author} {\bibfnamefont {A.}~\bibnamefont
  {R\"{u}ppert}}, \ and\ \bibinfo {author} {\bibfnamefont {A.}~\bibnamefont
  {Geiger}},\ }\href@noop {} {\bibfield  {journal} {\bibinfo  {journal} {Chem.
  Eur. J. of Chem. Phys.}\ }\textbf {\bibinfo {volume} {9}},\ \bibinfo {pages}
  {2737} (\bibinfo {year} {2008})}\BibitemShut {NoStop}%
\bibitem [{\citenamefont {Matsumoto}\ \emph {et~al.}(2002)\citenamefont
  {Matsumoto}, \citenamefont {Saito},\ and\ \citenamefont
  {Ohmine}}]{Ohmine2002}%
  \BibitemOpen
  \bibfield  {author} {\bibinfo {author} {\bibfnamefont {M.}~\bibnamefont
  {Matsumoto}}, \bibinfo {author} {\bibfnamefont {S.}~\bibnamefont {Saito}}, \
  and\ \bibinfo {author} {\bibfnamefont {I.}~\bibnamefont {Ohmine}},\
  }\href@noop {} {\bibfield  {journal} {\bibinfo  {journal} {Nature}\ }\textbf
  {\bibinfo {volume} {416}},\ \bibinfo {pages} {409} (\bibinfo {year}
  {2002})}\BibitemShut {NoStop}%
\bibitem [{\citenamefont {Fern{\'a}ndez}\ \emph {et~al.}(2006)\citenamefont
  {Fern{\'a}ndez}, \citenamefont {Abascal},\ and\ \citenamefont
  {Vega}}]{Fernandez2006}%
  \BibitemOpen
  \bibfield  {author} {\bibinfo {author} {\bibfnamefont {R.}~\bibnamefont
  {Fern{\'a}ndez}}, \bibinfo {author} {\bibfnamefont {J.}~\bibnamefont
  {Abascal}}, \ and\ \bibinfo {author} {\bibfnamefont {C.}~\bibnamefont
  {Vega}},\ }\href@noop {} {\bibfield  {journal} {\bibinfo  {journal} {J. Chem.
  Phys.}\ }\textbf {\bibinfo {volume} {124}},\ \bibinfo {pages} {144506}
  (\bibinfo {year} {2006})}\BibitemShut {NoStop}%
\bibitem [{\citenamefont {Sciortino}\ \emph {et~al.}(2011)\citenamefont
  {Sciortino}, \citenamefont {Saika-Voivod},\ and\ \citenamefont
  {Poole}}]{Sciortino2011}%
  \BibitemOpen
  \bibfield  {author} {\bibinfo {author} {\bibfnamefont {F.}~\bibnamefont
  {Sciortino}}, \bibinfo {author} {\bibfnamefont {I.}~\bibnamefont
  {Saika-Voivod}}, \ and\ \bibinfo {author} {\bibfnamefont {P.}~\bibnamefont
  {Poole}},\ }\href@noop {} {\bibfield  {journal} {\bibinfo  {journal} {Phys.
  Chem. Chem. Phys.}\ }\textbf {\bibinfo {volume} {13}},\ \bibinfo {pages}
  {19759} (\bibinfo {year} {2011})}\BibitemShut {NoStop}%
\bibitem [{\citenamefont {Errington}\ and\ \citenamefont
  {Debenedetti}(2001)}]{Errington2001}%
  \BibitemOpen
  \bibfield  {author} {\bibinfo {author} {\bibfnamefont {J.~R.}\ \bibnamefont
  {Errington}}\ and\ \bibinfo {author} {\bibfnamefont {P.~G.}\ \bibnamefont
  {Debenedetti}},\ }\href@noop {} {\bibfield  {journal} {\bibinfo  {journal}
  {Nature}\ }\textbf {\bibinfo {volume} {409}},\ \bibinfo {pages} {318}
  (\bibinfo {year} {2001})}\BibitemShut {NoStop}%
\bibitem [{\citenamefont {Kesselring}\ \emph {et~al.}(2011)\citenamefont
  {Kesselring}, \citenamefont {Franzese}, \citenamefont {Buldyrev},
  \citenamefont {Herrmann},\ and\ \citenamefont {Stanley}}]{Kesselring2011}%
  \BibitemOpen
  \bibfield  {author} {\bibinfo {author} {\bibfnamefont {T.}~\bibnamefont
  {Kesselring}}, \bibinfo {author} {\bibfnamefont {G.}~\bibnamefont
  {Franzese}}, \bibinfo {author} {\bibfnamefont {S.}~\bibnamefont {Buldyrev}},
  \bibinfo {author} {\bibfnamefont {H.}~\bibnamefont {Herrmann}}, \ and\
  \bibinfo {author} {\bibfnamefont {E.}~\bibnamefont {Stanley}},\ }\href@noop
  {} {\bibfield  {journal} {\bibinfo  {journal} {Arxiv preprint
  arXiv:1112.2186}\ } (\bibinfo {year} {2011})}\BibitemShut {NoStop}%
\bibitem [{\citenamefont {Steinhardt}\ \emph {et~al.}(1983)\citenamefont
  {Steinhardt}, \citenamefont {Nelson},\ and\ \citenamefont
  {Ronchetti}}]{Steinhardt1983}%
  \BibitemOpen
  \bibfield  {author} {\bibinfo {author} {\bibfnamefont {P.~J.}\ \bibnamefont
  {Steinhardt}}, \bibinfo {author} {\bibfnamefont {D.~R.}\ \bibnamefont
  {Nelson}}, \ and\ \bibinfo {author} {\bibfnamefont {M.}~\bibnamefont
  {Ronchetti}},\ }\href@noop {} {\bibfield  {journal} {\bibinfo  {journal}
  {Phys. Rev. B}\ }\textbf {\bibinfo {volume} {28}},\ \bibinfo {pages} {784}
  (\bibinfo {year} {1983})}\BibitemShut {NoStop}%
\bibitem [{\citenamefont {Molinero}\ and\ \citenamefont
  {Moore}(2009)}]{Molinero2009}%
  \BibitemOpen
  \bibfield  {author} {\bibinfo {author} {\bibfnamefont {V.}~\bibnamefont
  {Molinero}}\ and\ \bibinfo {author} {\bibfnamefont {E.~B.}\ \bibnamefont
  {Moore}},\ }\href@noop {} {\bibfield  {journal} {\bibinfo  {journal} {J.
  Phys. Chem. B}\ }\textbf {\bibinfo {volume} {113}},\ \bibinfo {pages} {4008}
  (\bibinfo {year} {2009})}\BibitemShut {NoStop}%
\bibitem [{\citenamefont {Laage}\ and\ \citenamefont
  {Hynes}(2006)}]{Laage2006}%
  \BibitemOpen
  \bibfield  {author} {\bibinfo {author} {\bibfnamefont {D.}~\bibnamefont
  {Laage}}\ and\ \bibinfo {author} {\bibfnamefont {J.}~\bibnamefont {Hynes}},\
  }\href@noop {} {\bibfield  {journal} {\bibinfo  {journal} {Science}\ }\textbf
  {\bibinfo {volume} {311}},\ \bibinfo {pages} {832} (\bibinfo {year}
  {2006})}\BibitemShut {NoStop}%
\end{thebibliography}

%

\end{document}